\documentclass[11pt,twoside]{article}
\usepackage{./asp2010}

\begin{document}

\title{OB stars in the Leading Arm of the Magellanic Stream}
\author{C. Moni Bidin,$^1$ D.~I. Casetti-Dinescu,$^2$, R.~A. M\'endez$^3$, T.~M. Girard$^2$,
K. Vieira$^4$, V.~I. Korchagin$^5$, and W.~F. van Altena$^2$
\affil{$^1$Instituto de Astronom\'ia, Universidad Cat\'olica del Norte, Av. Angamos 0610, Antofagasta, Chile}
\affil{$^2$Yale University, Dept. of Astronomy, P.O. Box 208101, New Haven, CT 06520-8101}
\affil{$^3$Universidad de Chile, Departamento de Astronom\'ia, Casilla 36-D, Santiago, Chile}
\affil{$^4$Centro de Investigaciones de Astronomia, Apartado Postal 264, M\'erida 5101-A, Venezuela}
\affil{$^5$Institute of Physics, Southern Federal University, Stachki 194, Rostov-on-Don 344090, Russia}
}

\begin{abstract}
We present our spectroscopic program aimed to study some new interesting features recently discovered in the
Magellanic Cloud System. These were revealed by the spatial distribution of OB-type candidate stars, selected
based on UV, optical and IR photometry and proper motions from existing large-area catalogs. As a pilot study
of our project, we are studying OB-star candidates in the Leading Arm (LA) of the Magellanic Stream, a gaseous
tidal structure with no stellar counterpart known so far. Our targets group in three clumps near regions of high
HI density in the LA. If confirmed, these young stars would evidence recent star formation in the LA, and they
would help better understand and constrain the formation of the LA and its interactions with the Milky Way.
\end{abstract}

\section{OB stars in the Magellanic Cloud System}
\label{s_general}

The Magellanic Cloud System comprises a complex ensemble of gaseous tidal structures originating from the
Magellanic Clouds, such as the extended Magellanic Stream (MS), encompassing $\sim$200~degrees across the sky
\citep{Nidever10}, the Bridge, and the Leading Arm (LA). The presence of these structures reveals the complex
gravitational interplay between the Clouds and the Milky Way, but their formation is still largely unclear.

In a recent study, \citet{Casetti12} listed 567 OB star candidates in a wide area including the periphery
of the Clouds, the Bridge, the LA, and part of the MS. These stars were selected from cuts in magnitude, colors,
and proper motions, after merging the UV photometry from the Galaxy Evolution Explorer survey
\citep[GALEX,][]{Bianchi11}, the IR data of the Two Micron All Sky Survey \citep[2MASS,][]{Skrutskie06}, the
optical photometry of the American Association of Variable Star Observers All Sky Photometric Survey
\citep[APASS,][]{Henden11}, and proper motions and photometry from the Southern Proper Motion Program~4
\citep[SPM4,][]{Girard11}. The spatial distribution of these candidates revealed various interesting features,
namely: 1) a well-populated wing of the Small Magellanic Cloud (SMC), continuing westward with two branches;
2) a narrow path from the SMC wing toward the Large Magellanic Cloud (LMC), offset from the ridge in the Bridge;
3) a well-populated periphery of the LMC; 4) a few scattered candidates in the MS and some overdensities in the
LA.

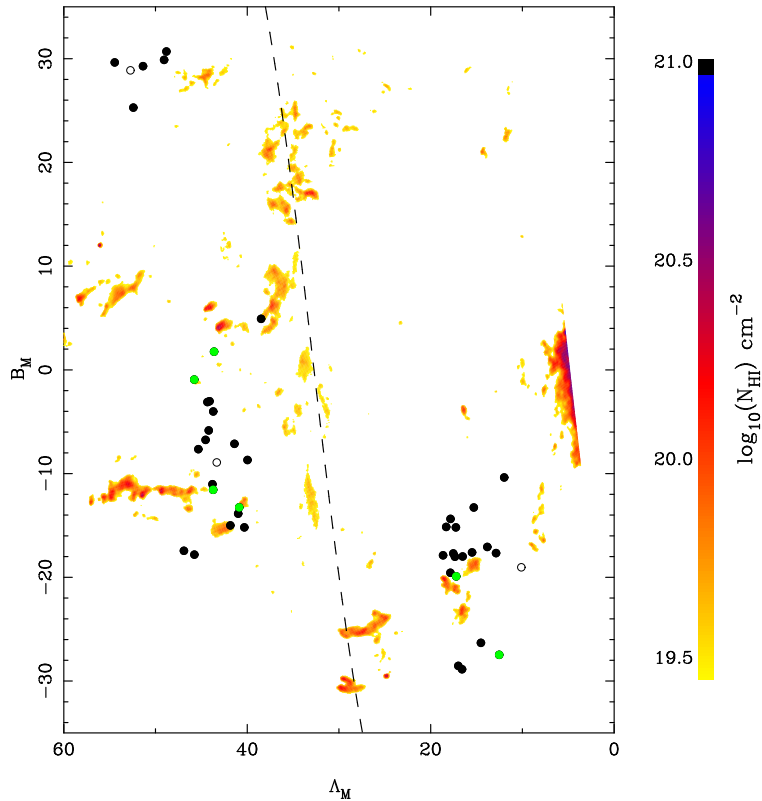
\articlefigure[angle=-90,width=10cm]{MoniBidin_f1.ps}{sp_distrib}{The spatial distribution in Magellanic coordinates
of our OB candidates (black circles). Filled black circles show the stars observed spectroscopically, filled green
circles show those candidates with velocities larger than 150~km~s$^{-1}$. The HI distribution for LSR velocities
between 150 and 400~km~s$^{-1}$ from the GASS survey \citep{Kalberla10} is also shown. The dashed line represents
the Galactic plane.}

We have recently started an extensive spectroscopic investigation, to study in detail the structures
outlined by \citet{Casetti12}. The first aim of our project is the collection of intermediate-resolution
spectra to derive the radial velocity (RV) and the stellar parameters of the OB candidates, to confirm their
membership to the Magellanic System and study their kinematics. Follow-up high-resolution studies, aimed at a
detailed chemical analysis, are planned for confirmed members of the Magellanic System.

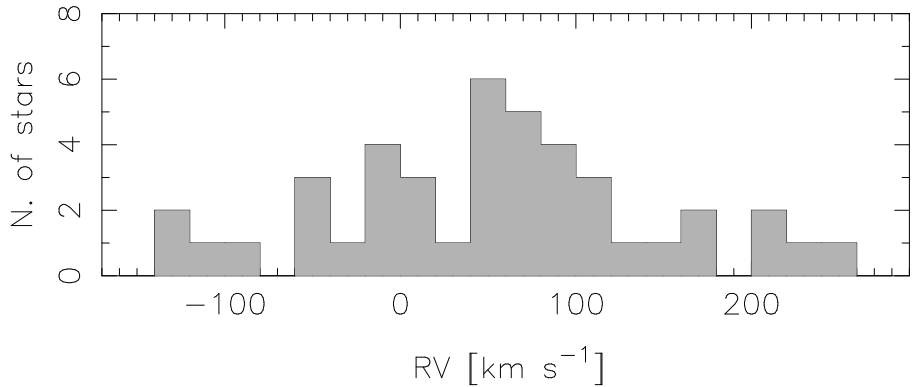
\articlefigure[angle=-90,width=12cm]{MoniBidin_f2.ps}{f_histo}{RV distribution of the observed stars.}

The magnitude and color selections of \citet{Casetti12} 
provided type O and B candidates; the proper-motion selection provided likely distant candidates.
Thus, the aim was to search for hot, young main sequence (MS) stars that are  too distant
to be Galactic stars.
Nevertheless, intrinsically fainter foreground white dwarfs (WDs) and subdwarf B- and
O-type stars (sdB's) can contaminate the sample. RVs can help identify the Magellanic System members because,
as indicated by the kinematics of the related gas \citep[Figure~3 of][]{Venzmer12}, high values in excess to
150~km~s$^{-1}$ are expected. Halo stars may have such high RVs, however, these stars are expected to be
early A type, or blue horizontal branch stars, thus redder than our color selections. 
Since, via proper motions, we eliminate nearby stars (i.e., WDs),
we believe our major source of contamination is from subdwarf O and B stars.  

The great majority of hot MS stars are found in binaries \citep{Sana12}, with preference
to equal-mass close systems. Close binaries are extremely common even among sdB's \citep{Maxted01}, although
they could be less frequent among older halo objects \citep{Moni08}. Hence, RVs are not conclusive to assess
the membership of the targets to the Magellanic System. The measurement of temperature and gravity can easily
identify WDs but, as shown by \citet{Salgado13}, these parameters cannot distinguish MS stars from sdB's
in the interval T$\approx$13\,000--17\,000~K. However, as discussed by the same authors, other indicators may be
used. For example, the surface helium abundance can be very indicative, because the atmosphere of sdB's in this
temperature range is depleted of helium by more almost a factor of 100 \citep{OToole08,Moni12} due to
gravitational settling \citep{Greenstein67,Baschek75}. The rotational velocity is also indicative, because fast
rotators are common among early-type MS stars, but not among sdB's \citep{Geier12}. However, given the large
difference in distance between the genuine members and the foreground contaminants, the spectroscopic mass is
probably the most efficient discriminant among them
\citep[see, e.g.,][for a successful analysis based on this criteria]{Moehler00}.

As a pilot study of our extensive program, we have recently started the analysis of the overdensities of OB star
candidates found by \citet{Casetti12} in the direction of the LA.
In Figure~\ref{sp_distrib}, we show the spatial distribution of our OB candidates in the LA region.
The HI distribution for velocities between 150 and 400~km~s$^{-1}$ is also shown, from the GASS 
survey \citep{Kalberla10}.

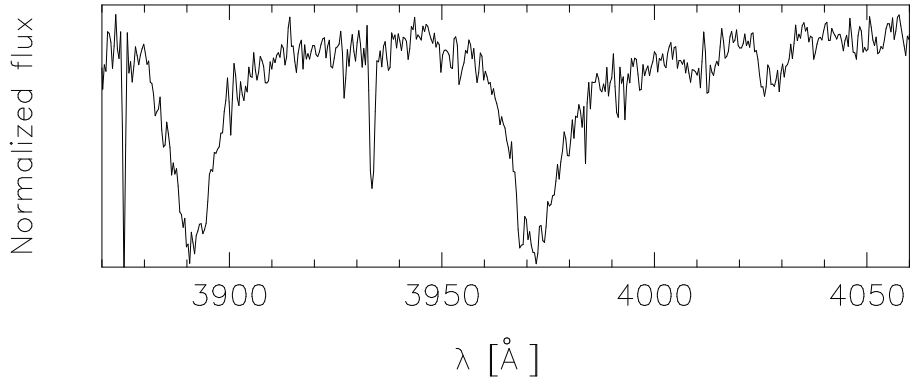
\articlefigure[angle=-90,width=12cm]{MoniBidin_f3.ps}{f_spec}{Spectrum of a LA hot star candidate from our sample.}

\section{Star formation in the Leading Arm}
\label{s_la}

The LA is a complex gaseous feature preceding the LMC in its orbit, whose origin is likely tidal, but 
its morphology can not be explained by purely tidal models \citep{Diaz11}. It comprises
four sub-structures \citep{Venzmer12}, with no known stellar counterpart. 
It has been suggested that the LA is hydrodynamically interacting
with both the gaseous Galactic disk \citep{McClure08} and the hot halo \citep{Diaz11}. The discovery of OB stars in
the LA would demonstrate that recent star formation occurred possibly as a consequence of these interactions, 
opening a new insight into the complex dynamical environment of the Magellanic System.

The spectra of forty-two OB star candidates from the \citet{Casetti12} sample in three LA overdensities above and below the
Galactic plane, were collected. This was done during two observing
nights with the IMACS spectrograph at the 6.5m Baade telescope at Las Campanas Observatory. The 1200 l/mm
grating at the f/4 camera was employed at first order, with a blaze angle of 17$\deg$ and a $0\farcs75$-wide slit,
for a resulting resolution of 1.3~\AA\ (R$\approx$3500) in the range 3650--5230~\AA. The average seeing during
observations was $0\farcs7$, and the resulting spectral signal-to-noise ratio was higher than 50 for all the stars.
The spectra were reduced and extracted with standard IRAF\footnote{IRAF is distributed by the National Optical
Astronomy Observatories, which are operated by the Association of Universities for Research in Astronomy, Inc.,
under cooperative agreement with the National Science Foundation.} routines.

Classical cross-correlation techniques \citep{Tonry79}, as implemented in the IRAF {\it fxcor} task, were employed
to measure RVs. In absence of a prior knowledge of the exact temperature and gravity of the targets, the synthetic
spectrum of a MS B-type star drawn from the library of \citet{Munari05} was adopted as template. However, a mismatch
between the parameters of the template and object spectra enhance the uncertainties but does not affect the results,
especially for hot stars \citep{Morse91,Moni11}. The RVs of three standard stars from \citet{Chubak12} and two
spectrophotometric standard hot stars from \citet{Hamuy94} were inspected to estimate and correct a zero-point
offset of $\approx$5~km~s$^{-1}$ in both nights. The final error, taking into account the relevant sources of
uncertainties (wavelength calibration, cross-correlation, and zero-point correction) resulted between 3 and
14~km~s$^{-1}$, depending on the star, but close to $\approx$5~km~s$^{-1}$ for most of the targets.

The RV distribution of the observed targets is shown in Figure~\ref{f_histo}. These preliminary results evidence
that the sample is highly contaminated by foreground Galactic stars, because most of the stars have
RV$<140$~km~s$^{-1}$, incompatible with the motion of the LA \citep{Venzmer12}. Nevertheless, eight stars are found
at RV$>140$~km~s$^{-1}$.
However, as discussed in Section~\ref{s_general}, RVs alone are not sufficient to establish LA
membership. The spectra are currently being fitted with the routines developed by
\citet{Bergeron92} and \citet{Saffer94}, as modified by \citet{Napiwotzki99}, to derive the stellar parameters
(temperature, gravity, and surface helium abundance) of the targets. In Figure~\ref{f_spec} we show the spectrum of
a very promising star, whose RV ($\sim$170~km~s$^{-1}$) excellently agrees with the expectation for a LA member,
and whose broad helium lines visible in the Figure point to a fast rotating MS star.

\acknowledgements This investigation is based on data gathered with the 6.5-meter Baade telescope, located at Las
Campanas Observatory, Chile (program ID: CN2013A-152).

\bibliography{MoniBidin2.bib}

\end{document}